**Assessing Pedestrian and Bicycle Treatments at Complex Continuous Flow Intersections**

Ishtiak Ahmed (Corresponding Author)[1], Shannon Warchol[2], Chris Cunningham[3], and

Nagui Rouphail[4]

[1]PhD Student, Department of Civil, Construction, and Environmental Engineering,

North Carolina State University, Campus Box 7908, Raleigh NC 27695-7908.

Email: iahmed2@ncsu.edu

[2]Senior Engineer, Kittelson & Associates, Inc., 272 N Front St #410, Wilmington, NC 28401

[3]Director, Highway System, Institute for Transportation Research and Education,

North Carolina State University, 909 Capability Drive, Raleigh NC 27695-7908

[4]Professor, Department of Civil, Construction, and Environmental Engineering,

North Carolina State University, Campus Box 7908, Raleigh NC 27606

**ABSTRACT**

This study evaluated the performance of pedestrian-bicycle crossing alternatives at Continuous

Flow Intersections (CFI). Further, a comparison was also performed of CFI crossing types

against a standard intersection designed to provide an equivalent volume-to-capacity ratio. Three

CFI crossing alternatives were tested, namely Traditional, Offset, and Midblock crossings. In total, 12 alternative scenarios were generated by incorporating two bicycle path types and two right-turn control types. These scenarios were analyzed through microsimulation on the basis of stopped delay and number of stops.

Simulation results revealed that the Offset crossing alternative incurred the least stopped delay for all user classes, including motorized traffic. The Traditional crossing generated the least number of stops for most route types. The Midblock crossing can be considered as a supplement to either the Offset or Traditional crossing depending on the specific origin-destination patterns at the intersection. The exclusive bicycle path performed better than the shared-use path in most cases. When compared with an equivalent standard intersection, aggregated results showed significant improvement for all CFI crossing types with respect to stopped delay, but the standard intersection had an equal or fewer number of stops for most routes investigated.  Regarding the effect on vehicular movement, the lowest volume-to-capacity ratio of the main intersection was incurred by the Offset crossing. Future research includes incorporating pedestrian-bicycle safety, comfort, and the relative effects of these crossing alternatives on additional vehicular performance measures.

## INTRODUCTION

Continuous Flow Intersections (CFI) are commonly used at the junction of two arterials with significant traffic volumes. Thus, they are often located in urban or suburban areas where pedestrian and bicycle generators are present. As such, efforts should be made to provide efficient safety and mobility for these non-motorized modes at CFIs. A number of crossing alternatives are currently being used, or have been proposed for CFIs, but research on the operational effectiveness of those crossing alternatives is lacking. This paper presents the tradeoffs of three crossing alternatives at CFIs in terms of stopped delay and the number of stops through microsimulation. It further compares the alternatives to a standard intersection designed for an equivalent volume-to-capacity ratio. The vehicular volume-to-capacity ratio at the main intersection was also estimated to evaluate the effect of different crossing options on those movements. While the focus of this effort is on the operational impacts of alternatives for pedestrians and bicyclists, practitioners should also consider the safety and user comfort aspects of the alternatives. The remainder of the paper lays out the current state of literature, the methodology used in this paper, the simulation results, and a summary of the major findings.

## LITERATURE REVIEW

The application of microsimulation tools to assess both vehicular and non-motorized users' safety and mobility at various types of intersections has been used in several studies. The concept of simulating pedestrians as vehicles in VISSIM was introduced by *(1)*. It also recalibrated the models to minimize earlier limitations leading to a successful calibration of pedestrians in VISSIM. As modeling relates to alternative intersection design choice from a research perspective, Holzem et al. *(2)* used VISSIM to test different pedestrian-bicycle crossing alternatives at superstreets.

Several studies demonstrated the assessment of vehicular operation at Diverging Diamond Interchanges (DDI) using VISSIM *(3–5)*.

The application of microsimulation in assessing CFI performance was found as early as 2005 *(6)*. That study demonstrated the assessment of one pedestrian crossing type at three different geometric designs of the CFI using VISSIM. The use of linear programming along with microsimulation was utilized in a few studies as well. A dynamic control strategy to reduce excess vehicular delay incurred by pedestrian crosswalks in a CFI was proposed by *(7)*. This study compared pedestrian safety and mobility for two types of crossing facilities in a four-legged CFI, namely Traditional and Offset, using VISSIM. The traditional crossing generated less crossing time than the offset crossing due to the straightforward structure but incurred additional delays to the vehicular movements. A multi-objective mixed-integer programming model was proposed by *(8)* to achieve the best operational performance of a CFI by changing the CFI type, configuration of the right-turn lane, distance to the displaced left-turn junction, and signal timing plan. However, it did not focus on the crosswalk geometries of the CFI. Zhao et al. *(9)* proposed to improve the operation of a CFI by shifting the crossing location of left-turning bicycles to the midblock location so there is no conflict with through traffic. A linear programming tool was used to optimize the geometry and signal timing and it was tested by simulating a real intersection in VISSIM.

Several studies proposed analytical frameworks to evaluate the operation of CFI and other alternative intersections in terms of pedestrian and bicycle crossings as well as vehicular movements. Wang et al. *(10)* developed an analytical model to calculate pedestrian delay at a CFI for three types of crossings and tested its accuracy with VISSIM. In addition to a traditional and an offset crossing, it demonstrated the application of an exclusive pedestrian phase, although it did not measure delay to vehicles accrued by pedestrians crossing such a large intersection footprint diagonally. In addition, the model requires the signal system to be fix-timed. FHWA published an analytical tool called "Cap-X" (*11*) that compares the performance of eight types of intersections including CFIs for different vehicle demands and lane configurations. Virginia Department of Transportation (VDOT) developed a tool (*12*) to analyze the performance of 26 Alternative Intersections and Interchanges in terms of vehicular congestion, safety, and pedestrian accommodation for screening purposes. Although these studies provide a quick sketch-level assessment of alternative intersections, the methods are deterministic in nature and cannot capture the stochasticity involved with the demand and capacity of the intersections

A few studies discussed the pedestrian-bicycle accommodations of the CFI using only qualitative assessment. Chlewicki (*13*) demonstrated different aspects of four types of crossing alternatives in a CFI, two of which are currently in practice. Recommendations were made regarding the placement of crossings, sidewalks, signal plans, median reductions, and separating turning movements. Several guideline reports and tools have been published regarding the mobility and safety of pedestrian-bicycles in a CFI. Among these, the reports by the Utah Department of Transportation and FHWA (*14–15*) discussed the configuration of pedestrian-bicycle crossing facilities in a CFI along with its signal timing plan. These provided general ideas regarding the current practices of various CFI installations both in terms of vehicular movements and pedestrian-bicycle crossing alternatives. The design considerations for signalized and unsignalized right turns along with discussions of the tradeoffs between single versus multi-staged crossings are also described.

To evaluate pedestrian-bicycle crossing facilities, the most common measures used in past studies are average or maximum delay per route per pedestrian (*2–3, 6*), average or total stops per pedestrian crossing (*2–3*), total stopped delay per pedestrian crossing geometry (*2–3*) and travel time per pedestrian crossing geometry (*2, 7*). Coates et al. (*7*) used exposure rate and time to cross to evaluate the safety and mobility of pedestrians at a CFI, respectively. The Highway Capacity Manual (*16*) proposes a Level of Service criteria based on average delay to evaluate pedestrian and bicycle crossing facilities. VJust (*12*) analyzes pedestrian accommodation using three performance measures – namely pedestrian safety, wayfinding or crosswalk alignment, and delay.

From the survey of available literature, it is apparent that several studies used microsimulation tools to evaluate pedestrian-bicycle crossing facilities at signalized intersections. However, only a few focused on CFIs and reported the performance on an aggregated level. Among those, research on testing different crossing alternatives and the variation in performance on a route-level was found to be lacking. The most common performance measures for pedestrian-bicycle mobility used by past studies are the descriptive statistics of stopped delay, number of stops, and travel time.

## METHODOLOGY

In this section, the three types of CFI pedestrian-bicycle crossing geometries considered in this study are described. Next, the development of alternative scenarios is explained by introducing two types of right turn control and bicycle paths. Finally, the model development process for an equivalent standard intersection is discussed.

### Pedestrian and Bicycle Crossing Geometries

Three types of crossing alternatives are proposed to be tested for both pedestrians and bicycles. Two of these alternatives – namely the Traditional and Offset crossing –are currently used in practice at existing CFIs. The third type, called the Midblock crossing, was proposed by (*13*); however, to the authors' best knowledge, is not currently in use at any CFI. In addition, two types of bicycle paths – namely shared-use paths and exclusive paths – are modeled in this study.

### Traditional Crossing

Illustration of a Traditional crossing for pedestrians and bicycles at a CFI is shown in Figure 1(a). This crossing configuration is widely used in the US. Similar to a standard four-legged intersection, the vehicular left-turn movement from one approach conflicts with the parallel pedestrian-bicycle crossing. Consequently, it requires four phases in the signal controller's ring-barrier system as the left turn and pedestrian-bicycle crossing cannot run simultaneously. However, the primary advantage of this crossing type is that all the users need only one stage to cross any leg. This crossing type was also termed a "Split 2-phase Crossing" in a past study *(7)*.

### Offset Crossing

Several CFIs in Mexico and a few in the US (e.g., East Eisenhower Blvd. & Madison Ave. in Loveland, CO and Beechmont Ave. & Five Mile Rd in Cincinnati, OH) have crosswalks aligned such that they do not conflict with the parallel left turns from the displaced left-turn legs. As shown in Figure 1(b), this design "offsets" the crosswalk toward the inside of the intersection, hence the term Offset crossing. As the left turn movement can simultaneously run with the parallel pedestrian-bicycle movement, this crossing geometry requires only two phases in the ring-barrier system. However, the major disadvantage of this crossing type is that pedestrians and bicycles need two phases to cross each leg of the intersection

### Midblock Crossing

This crossing type is similar to the Traditional crossing; however, the major street crossing is shifted to the "midblock" location from the main intersection, hence the term Midblock crossing. An advantage of this crossing is that it provides a very short travel path between the left corners of the NW and SW quadrant and between the right corners of NE and SE quadrant. Some routes, however, experience significant out of direction travel. In this setup, the vehicular signal timing can also be designed in such a way that the midblock crossing does not incur additional stops or delay to the vehicles as long as a median refuge is provided, as shown in Figure 1(c).

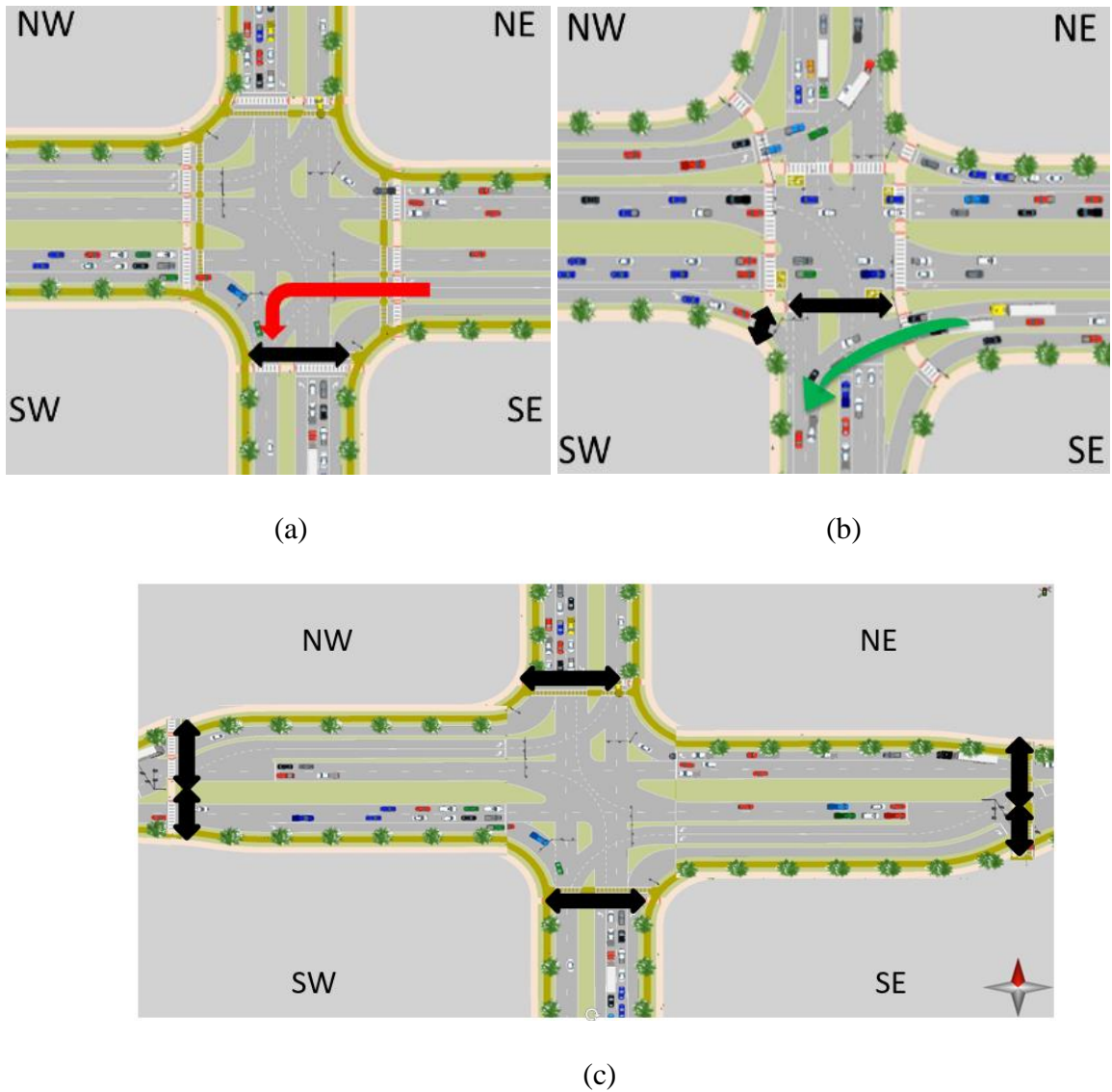

(a)                                              (b)

(c)

**Figure 1 Proposed CFI Crossing alternatives (a) Traditional (b) Offset (c) Midblock.**

*Courtesy: NCHRP 07-25 Research Team*

## Analysis Using Microsimulation

Microscopic simulation through PTV VISSIM 10.0 (*17*) was used to model the crossing alternatives of CFIs. The simulation run time was one hour, following a 15-minute warm-up period. Each treatment was replicated 25 times so that the results are statistically meaningful. The following paragraphs provide details of the analysis method using VISSIM.

*Base CFI Geometry Model*

The base model consisted of a four-legged CFI which is located in the middle of two standard signalized intersections in order to replicate a coordinated system in an urban corridor. Figure 2 shows a schematic of the entire model of all three intersections developed in VISSIM. The Offset crossing alternative is shown here. The major street (E-W direction) has three through lanes, two displaced left-turn lanes, and one channelized right turn lane on each approach. The displaced left-turn intersections are located 500 feet upstream of the main intersection. In this model, only the major street left-turn legs are displaced. The eastbound through movements are progressed through the three intersections. The minor street (N-S direction) approach has two through lanes, two standard left-turn lanes, and one channelized right turn lane.

*Pedestrian and Bicycle Model Construction*

Similar to many past studies (*1–2*), pedestrian-bicycles are modeled as "vehicles" in this experiment which allows interactions with vehicles. Sidewalks are modeled as "footpath" with a behavior type that allows the users to freely move without queueing, which is the default for vehicles. To ensure sufficient crossing samples, the input volumes at each of the eight origin points were 300 pedestrians and 300 bicycles per hour. Although not a realistic pedestrian volume for most intersections, allowing pedestrians to overtake one another freely provides realistic scenarios compared to field observations while shortening the needed run time for simulations.

Each quadrant has two origin points located 530 ft. away from the nearest corner of the main intersection along the major and minor street; therefore, each origin has seven destinations, each with a single route. The desired speed of pedestrians and bicycles is calibrated against field data from six standard intersections collected through prior research (*2*). Pedestrians are categorized into two groups – walkers and joggers – each consisting of 90% and 10% of the total pedestrian volume, respectively. The average speed of these two categories of pedestrians in the field dataset was 4.9 fps and 9.5 fps, respectively. Bicycle average speeds were set at 15.3 fps according to field observations.

*Scenario Generation*

The simulation models of the aforementioned crossing options were set up with various input variables. Among these, the three crossing geometries are described earlier. In total, 12 scenarios for bicycles and 12 scenarios for pedestrians were generated by combining additional variables with these crossing options. Table 1 and the following subsections discuss these variables. Note that the column "Bicycle path type" applies to bicycles only.

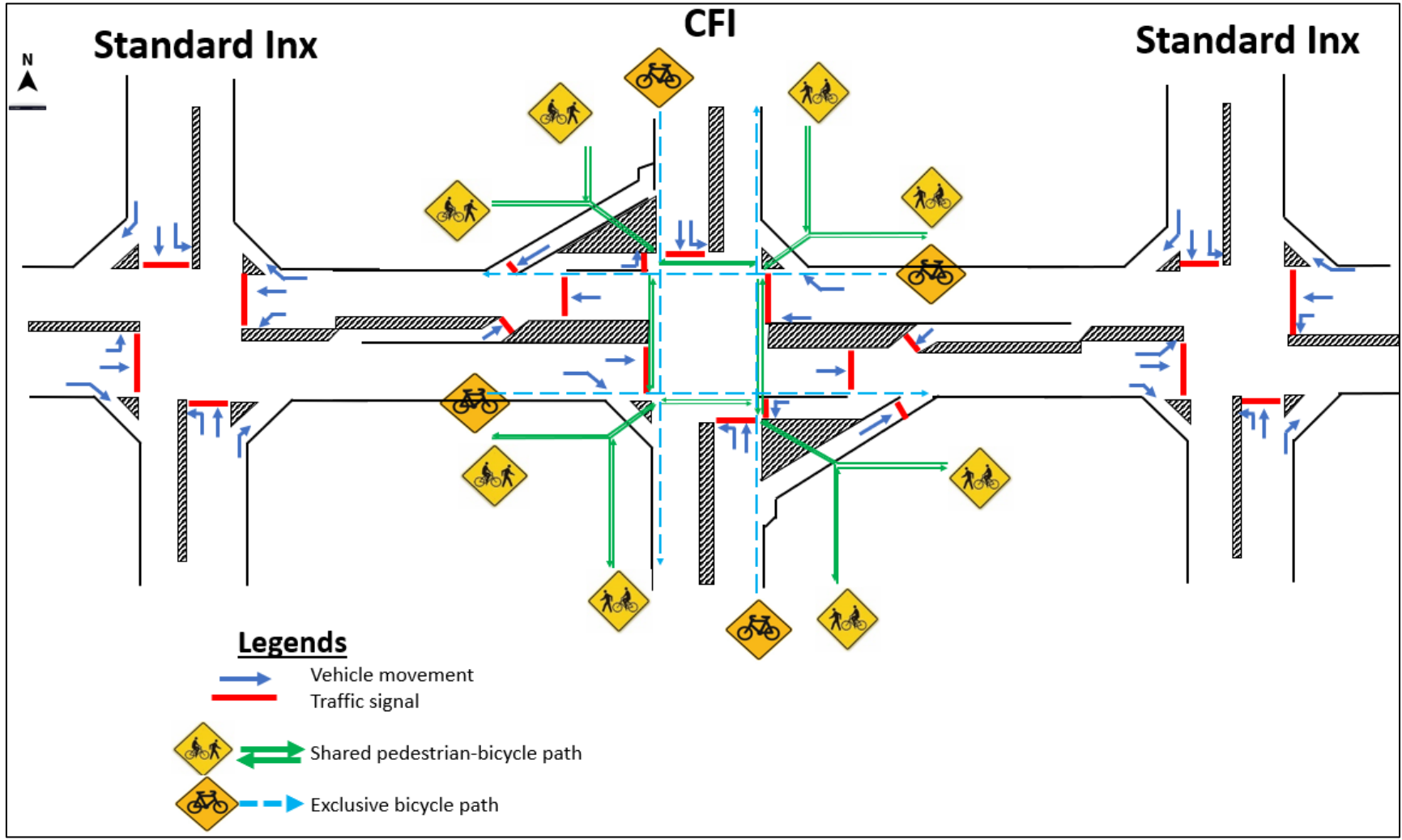

**Figure 2: Schematic of the CFI model with Offset crossing and neighboring intersections**

**TABLE 1 VISSIM Model Combinations Based on Variable Inputs**

| Combination no. | Bicycle path type | Right turn control | Crossing geometry |
|---|---|---|---|
| 1 | Exclusive | Signalized | Traditional |
| 2 | Shared | | |
| 3 | Exclusive | Unsignlaized | |
| 4 | Shared | | |
| 5 | Exclusive | Signalized | Midblock |
| 6 | Shared | | |
| 7 | Exclusive | Unsignlaized | |
| 8 | Shared | | |
| 9 | Exclusive | Signalized | Offset |
| 10 | Shared | | |
| 11 | Exclusive | Unsignlaized | |
| 12 | Shared | | |

**Bicycle Path Types:** Two types of bicycle paths were modeled: exclusive bicycle lane alongside the vehicular lane and shared-use path with pedestrians. The *exclusive bicycle lane* is a common cycle treatment in urban areas. It is a six-foot-wide lane adjacent to the rightmost vehicular lane and controlled by the vehicular signal at the intersection. In Figure 2, this path is shown by cyan arrows. The *shared-use path* is separated from traffic in a dedicated facility as shown by the green arrows in Figure 2. It is common particularly in locations with recreational cyclists and is typically found in suburban and urban areas. Since the operation of these two path types is different, it was essential to test both at a CFI. It should be noted that to be consistent with design practice, the shared-use paths are modeled as two-way paths, while the exclusive bicycle lanes are modeled as one-way causing some exclusive bicycle routes to be very long.

**Right Turn Control Types**: Two right turn-pedestrian interaction control types were modeled: yield-control and signal control. The control for pedestrian-bicycles crossing the right turn channelized lane is signalized only if the right turning vehicles are controlled by a signal as well.

Otherwise, ped-bicycles always have the priority to cross a channelized right turn lane through the use of a yield controlled crosswalk. This right-turn control type varies at different legs of a CFI as well as across different locations. Because both types are ubiquitous, both signalized and unsignalized right turns are modeled in VISSIM. Further, since pedestrian-bicycles do not always comply with the channelized right turn signal, a 50% compliance rate is assumed based on the outcomes from past studies on pedestrian-bicycle compliance rate (*18–19*). The priority rule in VISSIM enables the modeling of non-complying behavior as ped-bicycles cross the channelized right turn during red only if any vehicle is far enough (14 feet) from the right turn crosswalk.

*Traffic Volume*

Our target was to simulate a peak-hour condition during which the ped-bicycle delay is expected to be very high. On the other hand, an excessively high traffic volume would result in signal failure. Hence, a trial and error process was used using Cap-X (*11*) to select a volume for the given lane configuration such that the volume to capacity ratio (v/C) of any intersection remains in the range of 0.50 to 0.75. Based on that design, the following directional traffic volumes were obtained. The major street (east and westbound) of the CFI served 470 vehicles per hour (vph), 1,250 vph, and 200 vph for the left-turn, through, and right-turn movement respectively, in each direction. The minor street of the CFI served 310 vph, 880 vph, and 180 vph for the left-turn, through, and right-turn movement respectively, in each direction.

*Signal Timing*

All movements in the CFI are controlled using a single semi-actuated controller. For the given volume and lane configuration, the signal timing plan for the CFI intersection was developed by minimizing the cycle length while meeting the required green time so that the volume to capacity ratio (v/c) for any movement does not exceed 0.88. Figure 2 shows the ring-barrier diagram of 16 phases, their split times (colored green), and the movements in a CFI with Traditional crossing.

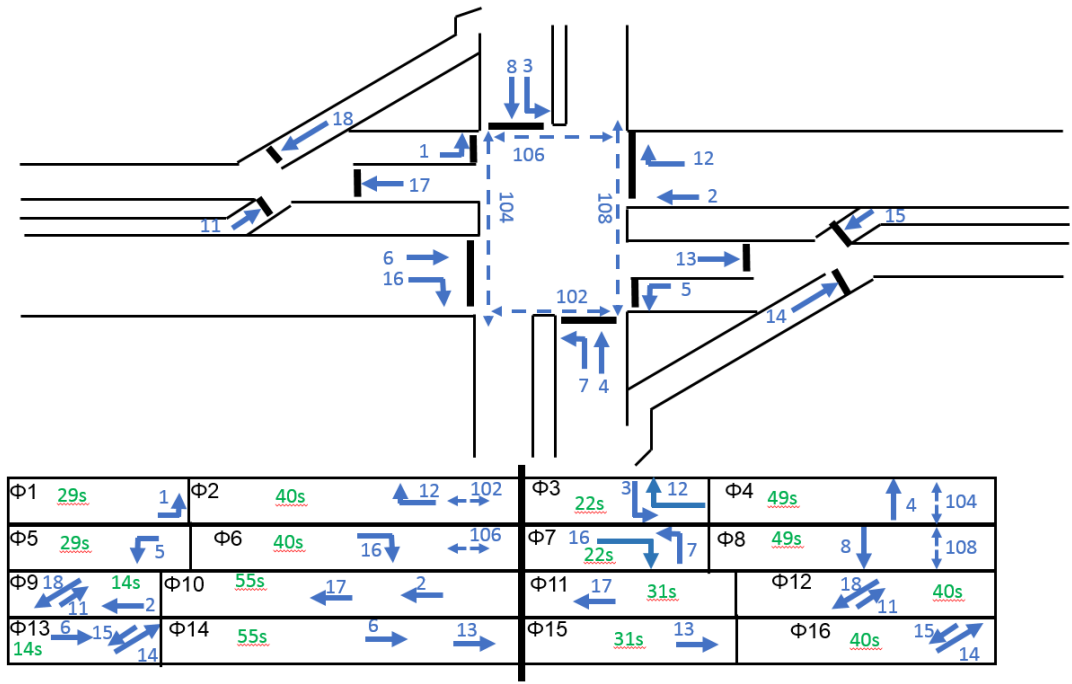

**Figure 3 Ring barrier diagram and movements in a CFI with Traditional crossing**

It should be noted that in the case of Traditional and Midblock crossing, the full advantage of installing a CFI – namely, two critical movement operation – cannot be achieved as the displaced left-turn movements conflict with the parallel pedestrian-bicycle movements. Therefore, these two crossing types require a longer cycle length than the Offset crossing. The cycle lengths obtained for Traditional and Midblock crossing were 140 seconds, while that for Offset crossing was 110 seconds.

**Equivalent Standard Intersection Modeling**

In order to contrast the performance of pedestrian-bicycle mobility between a CFI and a standard intersection, geometries of standard intersections equivalent to both Partial and Full CFI are obtained using Cap-X. Keeping the volume similar to that of the CFI models, different lane configurations of standard intersections were tested to achieve the same v/C of the main intersection. The standard intersection equivalent to the CFI consists of four through and two left-turn lanes on the major street, and three through and two left-turn lanes on the minor street, with one channelized right turn lane on each approach. The intersection v/C for the equivalent standard intersection was 0.65. The optimal cycle length for the standard intersection models generated by PTV VISTRO 7.0 was 135 sec.

**Comparison of Crossing Alternatives**

For the ease of discussion, pedestrian and bicycle routes are presented into four categories. As apparent from Figure 1(a), Diagonal (e.g., NE quadrant to SW quadrant), major street (e.g., NE quadrant to SE quadrant), minor street (e.g., NE quadrant to NW quadrant), and within the same quadrant are the most intuitive route types used to analyze any four-legged intersections. Here, the diagonal route type is further divided into two categories based on whether the users cross the DLT lanes (e.g., SE to NW quadrant in Offset crossing) or not (e.g., SW to NE quadrant in Offset crossing). The crossing alternatives are evaluated based on two performance measures for pedestrian and bicycles: stopped delay and number of stops. The measures obtained from the equivalent standard intersection model are also compared with their equivalent CFI model.

To estimate the effect of different crossing geometries on vehicular movements, the volume to capacity ratio (v/C) of the entire CFI for each crossing geometry is estimated according to the Highway Capacity Manual (*16*). Note that pedestrian and bicycle control type (yield and signal controls) is expected to affect the v/C of the intersection. However, in the comparison of the three crossing alternatives, this effect should be balanced out and hence, was not included in the v/C estimation process.

**RESULTS**

This section presents the results of the simulation runs. First, route-level variations of the crossing alternatives in terms of pedestrian and bicycle delay are discussed. Then, the impact of the crossing options on vehicular movements is described. Note that bicycles on the shared path generated--- for the most part—very similar performance measure as pedestrians and hence, its results are omitted from the figures. Also, results from unsignalized and signalized models are aggregated together for each crossing type.

**Performance of Pedestrian and Bicycle: Major Street Crossings**

Figure 4(a) and 4(b) show the average stopped delay and average number of stops, respectively, along the major street crossing. Performances of the three CFI crossing options along with an equivalent standard intersection crossing are shown for pedestrians and bicycles on the exclusive path. The error line on each bar represents ±1 standard error of the corresponding measure.

For both non-motorized users, the trend of stopped delay and number of stops are similar across the standard intersection and the three CFI crossing alternatives. The Midblock crossing generated the highest pedestrian stopped delays among the CFI crossing alternatives (82 seconds per pedestrian). The reason is that this crossing alternative not only has a higher number of phases than the Offset crossing but also requires two stages to cross the major street (see Figure 1(c)). The offset crossing generated the lowest stopped delay because of its shorter cycle length and fewer number of phases (54 seconds and 64 seconds per pedestrian and per exclusive bicycles, respectively). These benefits of the Offset crossing are attributed to the fewer number of conflicts due to the displaced left-turn legs in a CFI.

However, these displaced left-turn legs are also responsible for incurring the highest number of stops in the Offset crossing for both users (1.7 and 1.6 per pedestrian and per exclusive bicycles, respectively). Referring to Figure 1(b), the major street crossing is multi-stage in the Offset setting. The number of stops along this route is also very high in the Midblock crossing. Traditional crossing generated the lowest number of stops due to its simple and single-stage crossing design.

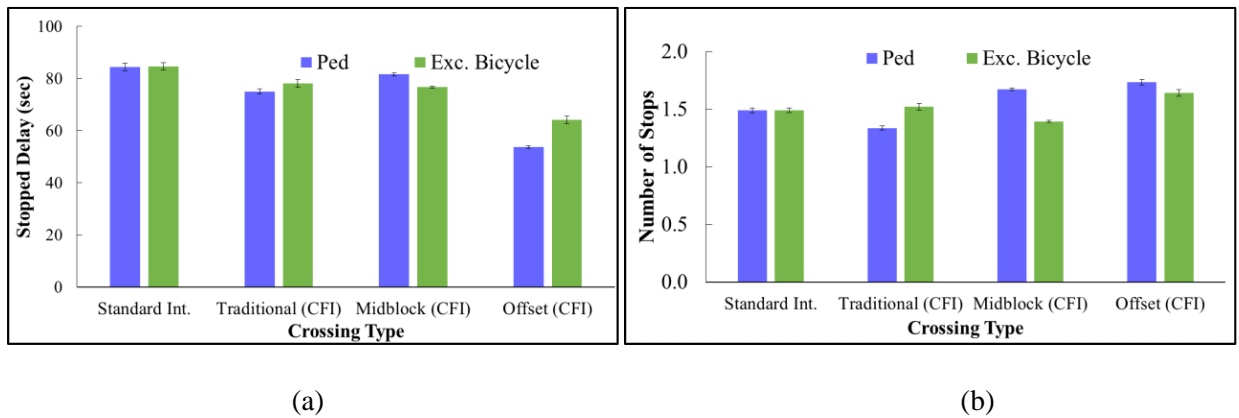

(a)                                                      (b)

**Figure 4 Performance of Major street crossing: (a) stopped delay (b) number of stops**

The standard intersection generated even higher stopped delay than the CFI Midblock crossing. Also, the number of stops it generated is higher than the CFIs with Traditional (for pedestrians) and Midblock crossing (for bicycles).

The performance for pedestrians and bicycles on the exclusive path was different for Offset and Midblock crossings. In the Offset setting, this is attributed to the fact that exclusive bicycles operate with vehicular movements that experience higher stopped delay than pedestrians but a lower number of stops. In the Midblock crossing, the difference is attributed to the ability to progress through multiple signals.

**Performance of Pedestrian and Bicycle: Minor Street Crossings**

Figure 5(a) and 5(b) show the average stopped delay and number of stops along the minor street crossing, respectively. The strongest contrast between Figure 4 and 5 is the difference between pedestrians' and exclusive bicycles' performance across different models. This difference is attributed to a combined effect of green time extension for the exclusive bicycles due to high vehicle demand on the east-west route and the conflict between pedestrians and right turn vehicles in the signalized setting. Another notable observation from Figure 5 is the high stopped delay and number of stops generated by the Offset crossing (82 seconds and 2.1 stops per pedestrian). Along this route, both Midblock and Traditional crossings at a CFI have similar and somewhat simpler configuration, while the Offset crossing is multi-stage. Therefore, both number of stops and stopped delay are highest in the Offset crossing along the minor street. The trend of Standard intersection crossing relative to the CFI crossing is similar when comparing the trends in Figure 4 and Figure 5.

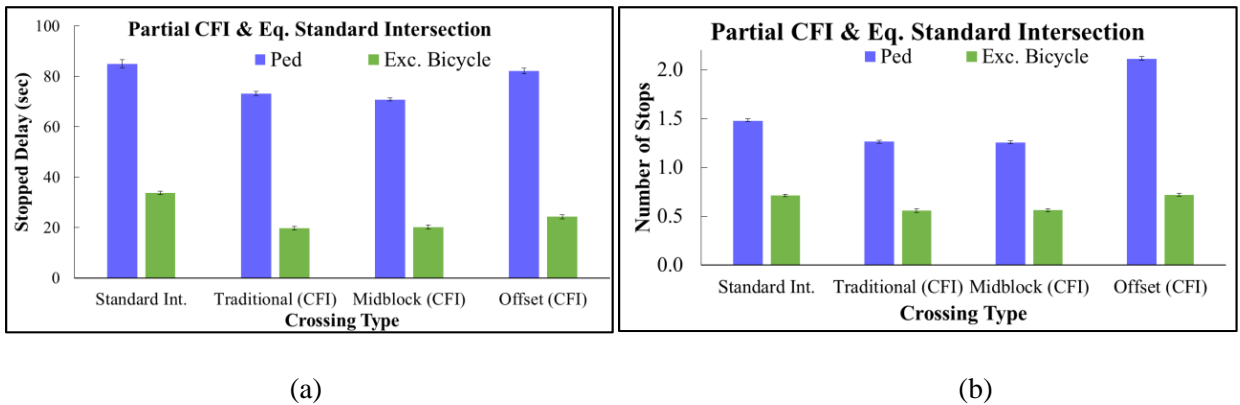

(a)                                                                                                    (b)

**Figure 5 Performance of Minor street crossing: (a) stopped delay (b) number of stops**

**Performance of Pedestrian and Bicycle: Diagonal Crossings without DLT legs**

Figure 6(a) and 6(b) show the average stopped delay and number of stops along the diagonal crossing without DLT legs, respectively. Note that both this route as well as the diagonal route with DLT legs are combinations of the major and minor street routes. Here, exclusive bicycles' delay and number of stops were consistently lower than those for pedestrians across all models. However, the differences are not large as for the minor street route. Midblock crossing generated the highest stopped delay per pedestrian (165 seconds) and number of stops (3.1 stops per pedestrian) primarily due to its long cycle length and multi-stage design of the major street crossing. Offset crossing generated the least stopped delay (92 seconds and 67 seconds per pedestrian and per exclusive bicycle, respectively). The Standard intersection generated a similar stopped delay and number of stops as the Traditional CFI crossing.

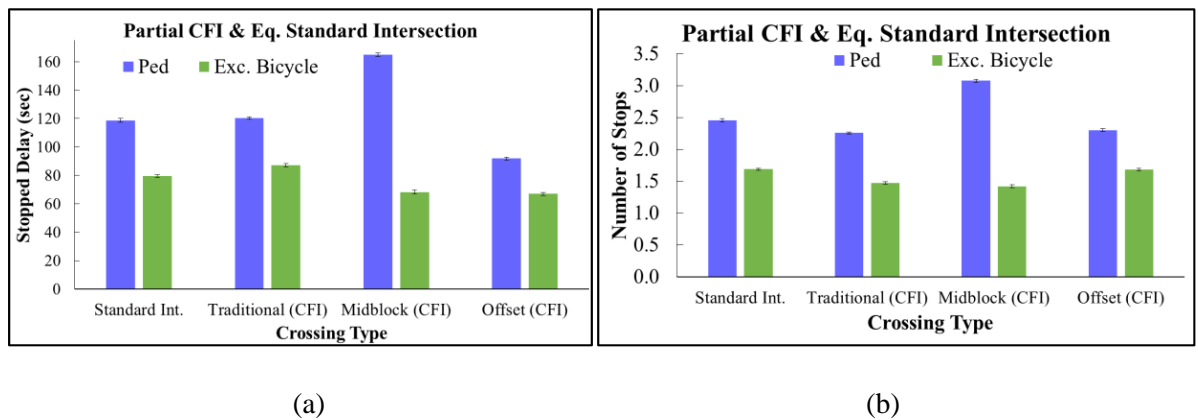

(a)                                                         (b)

**Figure 6: Performance of Diagonal crossing without DLT legs: (a) stopped delay (b) number of stops**

**Performance of Pedestrian and Bicycle: Diagonal Crossings with DLT legs**

Figure 7(a) and 7(b) show the average stopped delay and number of stops along the diagonal crossing with DLT legs, respectively. Note that this diagonal crossing type should exhibit a higher delay and higher number of stops than the diagonal crossing without DLT legs as it contains the highest number of signals for both users. However, as depicted in Figure 6 and Figure 7, the magnitudes of these differences are significant only for a few models. For instance, the Offset crossing generated a very high number of stops for pedestrians (3.5 per pedestrian). The average stopped delay experienced by the pedestrians in the Midblock crossing is also very high (175 seconds per pedestrian). Offset crossing generated the least stopped delay both per pedestrian (106 seconds) and per exclusive bicycle (49 seconds). The Midblock crossing generated the least number of stops for exclusive bicycles (1.4 per exclusive bicycle). On the other hand, Traditional crossing generated the least number of stops for pedestrians (2.3 per pedestrian). The relative performance of the Standard intersection along this route is similar to other routes. Its stopped delay (e.g., 123 seconds per pedestrian) and number of stops (e.g., 2.5 per pedestrian) lie in between the range of values obtained from the CFI crossing alternatives.

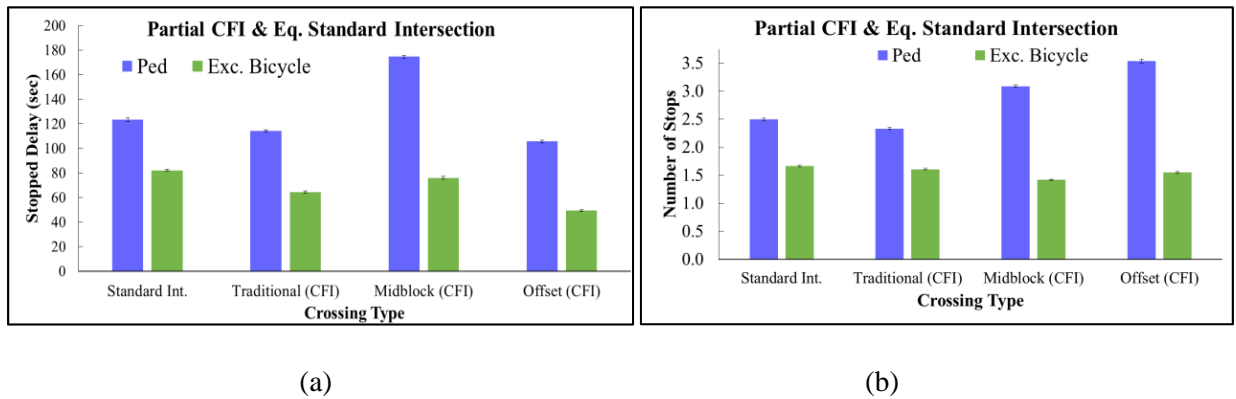

(a)                                          (b)

**Figure 7: Performance of Diagonal crossing with DLT legs: (a) stopped delay (b) number of stops**

Note that the small value of the standard error shown in Figure 4 through 7 indicates the number of simulation runs selected for this study was sufficient to evaluate the performance measures with appropriate confidence. It also indicates that the visual differences between the performances of the crossing alternatives are statistically significant.

**Performance of Vehicular Movements**

The traffic volumes in the simulation models were obtained from Cap-X to generate an intersection volume-to-capacity ratio in the range of 0.5 to 0.75. However, Cap-X estimates this volume-to-capacity ratio (v/C) based on the critical vehicular lane volume, which is based on intersection geometry and traffic volume data. It does not use the signal-timing information nor accounts for pedestrian and bicycle effects. Based on the actual signal timing setting, the v/C of the simulation output may vary significantly.

In this section, the v/C of the main intersection is used to compare the effect of different crossing alternatives on the vehicular movements. Table 2 shows the cycle length and intersection-wide v/C for the three CFI alternative models along with the Standard intersection model.

**Table 2: Intersection v/C of the models**

| Model name | Cycle length (sec) | v/C |
|:---:|:---:|:---:|
| Standard intersection | 135 | 0.81 |
| CFI with Traditional crossing | 140 | 0.67 |
| CFI with Midblock crossing | 140 | 0.67 |
| CFI with Offset crossing | 110 | 0.61 |

Note that the full advantage of a CFI is obtained only in the Offset crossing as the conflict between the through and opposing left-turn movement for vehicles is eliminated by the offset displaced left-turn legs. This is why the Offset crossing yielded the lowest cycle length among the three crossing alternatives. This also resulted in the lowest v/C of the intersection. On the other hand, the Traditional and Midblock crossing have similar signal plan, in which the cycle length and the resulting v/C are also the same. For the standard intersection, the geometry was selected using Cap-X to have a v/C within the range of 0.6 to 0.7. However, the resulting v/C for the selected cycle length and from the simulation output appeared to be very high.

## Summary

The analysis results provided a detailed assessment of the performances of different CFI crossing alternatives along with the standard intersection crossing. The performance of the vehicles in these alternative designs is also evaluated in terms of the intersection v/C for vehicles. These results indicate that (a) a Traditional crossing would generate the least number of stops for pedestrians and bicyclists along all routes; that (b) Midblock crossing would incur the longest stopped delay for most routes, and (c) that an Offset crossing would perform best in terms of stopped delay for most routes. Further, if adequate space is available, an exclusive bicycle path is preferable to the shared-use path as it would incur the lest stopped delay and number of stops in most cases. In terms of the vehicular movement degree of saturation, an Offset crossing would perform best.

Regarding the tradeoffs between a standard intersection and a CFI, a CFI would incur less stopped delay because of the reduced number of phases. A CFI would also incur less degree of saturation for vehicles than a standard intersection. However, a CFI with an Offset or a Midblock crosswalk would generate a higher number of stops than a standard intersection because of the increased number of stages.

## CONCLUSIONS

This paper investigated the tradeoffs between three crossing alternatives at CFIs in terms of stopped delay and number of stops. Trends were analyzed by categorizing the routes into four groups. Generally, the Offset crossing resulted in lower stopped delays but a greater number of stops. While the Offset required users to cross in multiple stages, the ability of users to cross concurrently with both through and left-turning traffic allowed for a higher green-to-cycle length ratio. The Midblock crossing can be considered as a supplement to either the Offset or Traditional crossing depending on the specific origin-destination patterns present at the intersection of interest.

A comparison was also performed of CFI crossing types to a standard intersection designed for an equivalent volume-to-capacity ratio. Aggregate results showed significant improvement for most CFI crossing types with respect to stopped delay, but the standard intersection had equal or fewer number of stops in most cases.

Local preference and dominant user type will likely dictate which performance measures are of most importance for any specific project. Future work, expected as part of NCHRP Project 07-25 (*21*) is expected to provide additional insight for practitioners seeking to balance operations with safety and user comfort. Additional operational analyses should consider methods to provide simultaneous progression for bicycles, pedestrians, and vehicles between Midblock crossings and those at the main intersection. Additionally, an exploration of how performance measures for any one specific crossing type varies with cycle length could provide more signal timing guidance to practitioners.

**DATA AVAILABILITY STATEMENT**

- Some or all data, models, or code that support the findings of this study are available from the corresponding author upon reasonable request.

- Some or all data, models, or code generated or used during the study are proprietary or confidential in nature and may only be provided with restrictions.

**ACKNOWLEDGMENTS**

The authors are grateful for the financial support of the Southeastern Transportation Research, Innovation, Development and Education (STRIDE) Center, with funding by the US DOT which sponsored this research. The opinions and conclusions stated in this paper strictly reflect those of the authors, and not of STRIDE, the US DOT, or its constituent members. The authors are also grateful to the NCHRP 07-25 research team for their support in this study.

## AUTHOR CONTRIBUTIONS

The authors confirm contribution to the paper as follows: study conception and design: C. Cunningham, S. Warchol, N. Rouphail, I. Ahmed; data collection: I. Ahmed, S. Gadiparthi; analysis and interpretation of results: N. Rouphail, C. Cunningham, S. Warchol,  I. Ahmed; draft manuscript preparation: C. Cunningham, S. Warchol, N. Rouphail, I. Ahmed, S. Gadiparthi. All authors reviewed the results and approved the final version of the manuscript.

**REFERENCES**

1.  Ishaque, M. M., and R. B. Noland. Pedestrian and Vehicle Flow Calibration in Multimodal Traffic Microsimulation. *Journal of Transportation Engineering*, 2009. 135: https://doi.org/10.1061/(ASCE)0733-947X(2009)135:6(338).

2.  Holzem, A. M., J. E. Hummer, C. M. Cunningham, S. W. O'Brien, B. J. Schroeder, and K. Salamati. Pedestrian and Bicyclist Accommodations and Crossings on Superstreets. Transportation Research Record: Journal of the Transportation Research Board, 2015. https://doi.org/10.3141/2486-05.

3.  Edara, P. K., J. G. Bared, R. Jagannathan. Diverging Diamond Interchange and Double Crossover Intersection-Vehicle and Pedestrian Performance. Presented at 3rd International Symposium on Highway Geometric Design, Chicago, IL, 2005.

4.  Schroeder, B. J., K. Salamati, and J. Hummer. Calibration and Field Validation of Four Double-Crossover Diamond Interchanges in VISSIM Microsimulation. Transportation Research Record: Journal of the Transportation Research Board, 2014. https://doi.org/10.3141/2404-06.

5.  Warchol, S., T. Chase, and C. Cunningham. Use of Microsimulation to Evaluate Signal-Phasing Schemes at Diverging Diamond Interchanges. Transportation Research Record: Journal of the Transportation Research Board, 2017. https://doi.org/10.3141/2620-02.

6.  Jagannathan, R., and J. G. Bared. Design and Performance Analysis of Pedestrian Crossing Facilities for Continuous Flow Intersections. Transportation Research Record: Journal of the Transportation Research Board, 2005. 1939. https://doi.org/10.1177/0361198105193900116.

7.  Coates, A., P. Yi, P. Liu, and X. Ma. Geometric and Operational Improvements at

Continuous Flow Intersections to Enhance Pedestrian Safety. Transportation Research Record: Journal of the Transportation Research Board, 2015. https://doi.org/10.3141/2436-07.

8.    Zhao, J., W. Ma, K. Head, X. Yang. Optimal Operation of Displaced Left-Turn Intersections: A Lane-Based Approach. *Transportation Research Part C: Emerging Technologies,* 2015. 61: 29-48.

9.    Zhao, J., X. Gao, and V. L. Knoop. An Innovative Design for Left Turn Bicycles at Continuous Flow Intersections. *Transportmetrica B: Transport Dynamics*, 2019. https://doi.org/10.1080/21680566.2019.1614496.

10.    Wang, T., J. Zhao, and C. Li. Pedestrian Delay Model for Continuous Flow Intersections under Three Design Patterns. *Mathematical Problems in Engineering*, 2019. https://doi.org/10.1155/2019/1016261.

11.    Lochrane, T., J. Bared. Capacity Analysis for Planning of Junctions (CAP-X). FHWA, U.S. Department of Transportation, 2011.

12.    Virginia Department of Transportation. Innovative Intersections and Interchanges. http://www.virginiadot.org/innovativeintersections/. Accessed July 31, 2019.

13.    Chlewicki, G. Improving pedestrian operations at innovative geometric designs. Presented at 5th Urban Street Symposium in North Carolina, 2017.

14.    Utah Department of Transportation. CFI Guideline: A UDOT Guide to Continuous Flow Intersections, July 2013. https://www.udot.utah.gov/main/uconowner.gf?n=10114119157568379. Accessed July 31, 2019.

15.    Steyn, H., Bugg, Z., Ray, B., Daleiden, A., Jenior, P., & J. Knudsen. *Displaced left turn*

*intersection: informational guide.* Publication FHWA-SA-14-068. FHWA, U.S. Department of Transportation, 2014.

16.    Transportation Research Board. *Highway Capacity Manual sixth edition: A guide for multimodal mobility analysis (6th Edition)*, 2016.

17.    PTV Group. PTV Vissim. http://vision-traffic.ptvgroup.com/nl/products/ptv-vissim/. Accessed July 31, 2019.

18.    Ren, G., Z. Zhou, W. Wang, Y. Zhang, & W. Wang. Crossing behaviors of pedestrians at signalized intersections: observational study and survey in China. Transportation Research Record: Journal of the Transportation Research Board, 2011. https://doi.org/10.3141/2264-08.

19.    Hummer, J. E. An Update on Superstreet Implementation and Research. Presented at 8th National Conference on Access Management, Transportation Research Board, Baltimore, MD., 2008.

20.    Williams, L. J. Tukey's Honestly Signiflcant Diflerence (HSD) Test. *Encyclopedia of Research Design,* 2010. Thousand Oaks, CA: Sage, 1-5.

21.    National Cooperative Highway Research Program. Guide for Pedestrian and Bicycle Safety at Alternative Intersections and Interchanges. https://apps.trb.org/cmsfeed/TRBNetProjectDisplay.asp?ProjectID=4183. Accessed July 31, 2019.